\documentstyle[12pt,a4wide]{article}
\begin{document}
\input epsf
\newcommand{\be}{\begin{equation}}
\newcommand{\ee}{\end{equation}}
\newcommand{\pr}{\partial}
\newcommand{\ie}{{\it ie }}
\font\mybb=msbm10 at 11pt
\font\mybbb=msbm10 at 17pt
\def\bb#1{\hbox{\mybb#1}}
\def\bbb#1{\hbox{\mybbb#1}}
\def\Z {\bb{Z}}
\def\R {\bb{R}}
\def\E {\bb{E}}
\def\T {\bb{T}}
\def\C {\bb{C}}
\renewcommand{\theequation}{\arabic{section}.\arabic{equation}}
\newcommand{\news}{\setcounter{equation}{0}}
\newcommand{\bx}{{\bf x}}

\title{\vskip -70pt
\begin{flushright}
\end{flushright}\vskip 50pt
{\bf \large \bf DOMAIN WALL NETWORKS ON SOLITONS}\\[30pt]
\author{Paul Sutcliffe
\\[10pt]
\\{\normalsize   {\sl Institute of Mathematics,}}
\\{\normalsize {\sl University of Kent,}}\\
{\normalsize {\sl Canterbury, CT2 7NZ, U.K.}}\\
{\normalsize{\sl Email : P.M.Sutcliffe@kent.ac.uk}}\\}}
\date{May 2003}
\maketitle

\begin{abstract}
\noindent 
Domain wall networks on the surface of a soliton are studied in
a simple theory. It consists of two complex scalar fields, 
 in (3+1)-dimensions, with a global $U(1)\times \Z_n$ symmetry, where $n>2.$
Solutions are computed numerically in which one of the fields forms
a Q-ball and the other field forms a network of 
domain walls localized on the surface of the Q-ball. 
Examples are presented in which the domain walls lie along the edges
 of a spherical polyhedron, forming junctions at its vertices.
It is explained why only a small restricted class of polyhedra can arise
as domain wall networks.

\end{abstract}
\newpage

\section{Introduction}\news\label{sec-intro}

\quad\, Any theory with multiple isolated vacua supports domain walls
that separate spatial regions containing different vacua.
Indeed domain walls arise in a number of diverse areas of particle physics,
cosmology and condensed matter physics.
Recently, there has been a renewed interest in the study of domain
walls due to their appearance in supersymmetric QCD \cite{DS} and 
their relation to D-branes \cite{Wi1}. More generally, interesting analogies
between domain walls and D-branes have been observed, with the 
interactions between domain walls providing insight into 
more complicated D-brane systems \cite{DV}. 
This provides further motivation for the study of configurations with
multiple domain walls, containing junctions and networks.

Domain wall junctions exist in theories with three or more vacua
and can be combined to form a planar network \cite{GT,CHT,Sa,BB1,BB2}.
It has been suggested in refs.\cite{BB3,BB4} that a three-dimensional 
network of domain walls might exist in which a spherical domain wall
is a host for a planar network confined to its surface, with the
spherical domain wall being stabilized by a fermi gas pressure,
 as in soliton stars \cite{Lee}.
According to refs.\cite{BB3,BB4} such networks could then form 
configurations resembling the Platonic solids and fullerene polyhedra,
 such as the truncated icosahedron of the buckyball. 
In this paper we explore these ideas, using a simple theory
in which a domain wall network 
is trapped on the surface of a Q-ball. 
We compute 
numerical solutions of the theory, presenting explicit examples
where a network of domain walls lies  
along the edges of a spherical polyhedron, forming junctions
at its vertices. However, only a  small restricted class
of polyhedra can arise in this why, as we explain using 
several examples.
Only two of the five Platonic solids (the cube and octahedron)
can be realized and no fullerene polyhedra can be obtained.
The argument which rules out most polyhedra is based on an examination
of the compatibility conditions which arise when assigning vacua
to the faces of 
the polyhedron, and its implications for the distribution
of domain wall tensions, which are required to be in equilibrium.
 The analysis
 does not depend on the type of soliton on which
the domain walls are trapped, and so applies generally to networks on
Q-balls, spherical domain walls, soliton stars, monopoles, Skyrmions etc.

\section{A Model with Q-balls and Walls}\news\label{sec-model}

\quad\, We consider a theory, in (3+1)-dimensions, with two complex
scalar fields $\psi,\phi$ and Lagrangian density
\be
{\cal L}=\frac{1}{2}\partial_\mu\psi\partial^\mu\bar\psi
+\frac{1}{2}\partial_\mu\phi\partial^\mu\bar\phi
-|\psi|^2\left(1+(1-|\psi|^2)^2\right)-\beta^2\left|(1-|\psi|^2)4|\psi|^2
-\phi^n/v^n\right|^2
\label{lag}
\ee
where $\beta$ and $v$ are real positive parameters and $n>2$ is an integer.
There is a global $U(1)\times \Z_n$ symmetry, corresponding to  
the transformations $\psi\mapsto e^{i\alpha}\psi$ and $\phi\mapsto 
\omega_n\phi$
where $\alpha$ is an arbitrary real parameter and $\omega_n=e^{2\pi i/n}.$ 
There is a conserved Noether charge associated with the $U(1)$ symmetry,
given by
\be 
Q={1\over 2i}\int\left(\bar\psi\partial_t\psi-\psi\partial_t\bar\psi\right)
d^3x
\label{q}\ee
and this allows the possibility of Q-ball solitons \cite{Co}.

The particular model (\ref{lag}) is chosen as one of the simplest examples 
supporting both solitons and domain walls, as we now explain in detail,
but there are many other theories which share these properties, with
various kinds of solitons playing host to the walls.

The field equations which follow from (\ref{lag}) have $\phi=0$ as a
solution, so a consistent way to find a family of solutions is to 
consider this reduction directly
in the Lagrangian density, to obtain the reduced expression
\be
{\cal L}_0=\frac{1}{2}\partial_\mu\psi\partial^\mu\bar\psi-U(|\psi|)
\label{lagqb}
\ee
where the potential function is given by
\be
U(f)=f^2(1+(1-f^2)^2)+16\beta^2f^4(1-f^2)^2.
\label{pot}
\ee
If $\beta=0$ then this is precisely the model 
in which Q-balls are studied in ref.\cite{BS}, to which we refer
the reader for further details. For small values of $\beta$ ie.
$\beta^2\ll 1,$ the final term in (\ref{pot}) may be treated as
a small perturbation of the $\beta=0$ theory, and the
Q-ball solutions will be almost unchanged.

A stationary Q-ball  has the form
\be
\psi=e^{i\nu t}f(r)\,,
\label{qbform}\ee
where  $f(r)$ is a real
radial profile function which satisfies the ordinary differential equation
\be
\frac{d^2f}{dr^2}=-\frac{2}{r}\frac{df}{dr}-\nu^2 f+U'(f)\,,
\label{profile}\ee
with the boundary conditions that $f(\infty)=0$ and $\frac{df}{dr}(0)=0.$

This equation can be interpreted as describing the motion of a
point particle moving with friction in a potential 
 $U_{\rm eff}(f)=\nu^2f^2/2-U(f)$. This leads to constraints \cite{Co}
on the potential $U(f)$ and the frequency $\nu$ in order for a 
Q-ball solution to
exist. Firstly, the effective mass of $f$ must be negative. 
Defining  $U''(0)=\nu_+^2>0$, then this implies that
$\nu<\nu_+$. For the potential (\ref{pot}) we have that $\nu_+=2$,
independent of $\beta.$
Next the minimum of $U(f)/f^2$ must be
attained at some positive value of $f$, say $0<f_0<\infty,$ and existence of
the solution requires that $\nu>\nu_-$ where 
\be \nu_-^2=2U(f_0)/f_0^2\,.
\label{range}\ee
For the potential (\ref{pot}) the minimum of $U(f)/f^2$ occurs at
$f=f_0=1,$ and this yields the value $\nu_-^2=2U(f_0)/f_0^2=2,$ 
again independent of $\beta.$
Hence, Q-ball solutions exist for the range $\sqrt{2}<\nu<2.$ 
\begin{figure}[ht]
\begin{center}
\leavevmode
\vskip -3cm
\epsfxsize=15cm\epsffile{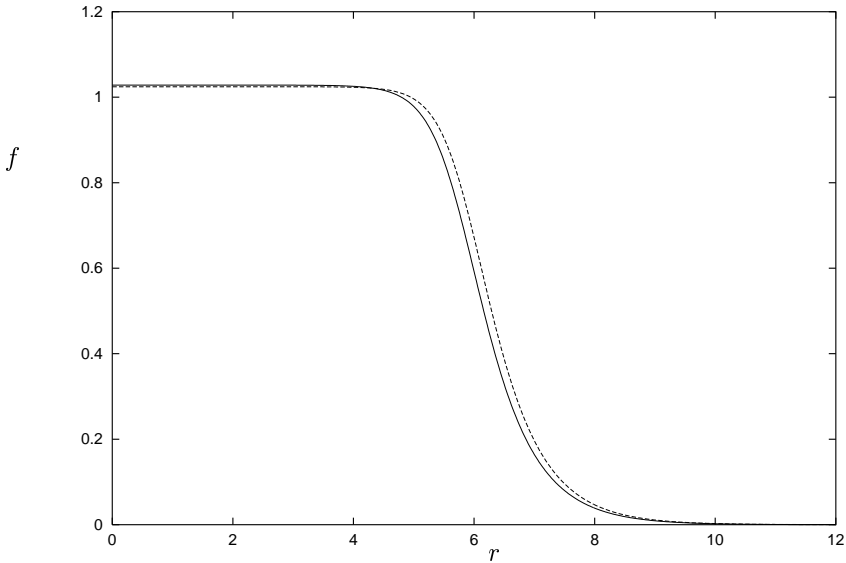}
\vskip -10.5cm
\caption{The Q-ball profile function $f(r)$ for $\nu=1.5$
with $\beta=0$ (solid curve) and $\beta=0.1$ (dashed curve).}
\label{fig-pro}
\end{center}
\end{figure}
In this paper we are concerned with large Q-balls, that is, 
$Q\gg 1,$ which corresponds to a frequency approaching the lower
limit $\nu_-=\sqrt{2}.$ Such large Q-balls are well described by
the thin wall limit, where  the profile function can be
approximated by a smoothed-out step function, with $f\approx 1$
inside the Q-ball and $f\approx 0$ outside, with a fairly sharp transition
at the surface of the Q-ball. This is demonstrated in fig.~\ref{fig-pro}
where we plot the profile function $f(r)$ (solid curve) for the
parameter values $\nu=1.5$ and $\beta=0.$ This profile function
was obtained by solving equation (\ref{profile}) numerically using
a shooting method. Also displayed in fig.~\ref{fig-pro} is the
profile function (dashed curve) for the parameter values $\nu=1.5$
and $\beta=0.1.$ This demonstrates that the shape of the Q-ball
 profile function is not very sensitive to the value of $\beta.$  
The Noether charge (\ref{q}) is slightly larger for
$\beta=0.1$ compared to $\beta=0,$ and   
in fact the main consequence of a non-zero value of $\beta$ is
a slight modification to the relationship between the charge $Q$ and
the frequency $\nu.$ 

Let us now turn our attention to solutions in which $\phi$ also 
has a non-trivial spatial dependence. The form of the Lagrangian (\ref{lag})
implies that for sufficiently small $\beta,$ ie. $\beta^2\ll 1$, the
final term in (\ref{lag}) will have a negligible effect
on the Q-ball solution and can be ignored. As we have seen in
 fig.~\ref{fig-pro} the value $\beta=0.1$ is within this regime, so
we shall make this choice from now on. In fact, one expects that the
effect of the final term in (\ref{lag}) will be even less for a
non-trivial $\phi$ field than in the case where $\phi=0,$ since the
additional contribution drives the $\phi$ field to minimize this term.

For simplicity we can therefore ignore the deformation of the Q-ball
and consider the $\phi$ field in the background of a fixed Q-ball 
solution. Since we are interested in static solutions for $\phi$
this reduces to studying the theory defined by the energy density
\be
{\cal E}=
\frac{1}{2}\partial_i\phi\partial_i\bar\phi
+\beta^2\left|4f^2(1-f^2)-\phi^n/v^n\right|^2
\label{energy}
\ee
where the index $i$ runs over the three spatial coordinates and 
$f(r)$ is a fixed Q-ball profile function. We shall choose
the Q-ball with $\nu=1.5,$ so that $f(r)$ is the solid curve
plotted in fig.~\ref{fig-pro}.

The expression $4f^2(1-f^2)$ is close to zero everywhere except
near the surface of the Q-ball, taking its maximal value 1, on
the surface $f=\frac{1}{\sqrt{2}}.$
 From fig.~\ref{fig-pro} we see that this surface
is a sphere of radius $r=r_*\approx 6.$
Far from the surface of the Q-ball we will
therefore find that $\phi\approx 0.$ However, on the surface of the
Q-ball the theory will resemble the Wess-Zumino model
\be
{\cal E}_{\rm WZ}=
\frac{1}{2}\partial_i\phi\partial_i\bar\phi
+\beta^2\left|1-\phi^n/v^n\right|^2
\label{WZ}
\ee
where, for the moment, we ignore the curvature of the sphere and
regard this energy as defining a two-dimensional theory in the plane.

The theory (\ref{WZ}) is the one studied in refs.\cite{GT,CHT,Sa} where
BPS junctions are investigated together with almost-BPS tilings of the plane.
We recall the salient features briefly. 
The theory has $n$ vacua, with $\phi$ proportional to the
$n$th roots of unity, ie.  $\phi=v\omega_n^j,$ where $j=1,..,n.$
In the following, for brevity, we shall sometimes refer to a vacuum by
its label $j.$  There are $n(n-1)/2$ different species of 
domain walls, each one connecting
any two distinct vacua, say $j$ and $k,$ with a tension
 (energy per unit length)
given by
\be
T=2^{3/2}\frac{n\beta v}{n+1}|\omega_n^{j}-\omega_n^{k}|.
\label{tension}
\ee
Thus the tension of a domain wall is proportional to the distance
in $\phi$ space between the two vacua it interpolates between.
These domain walls have a width of order $v/\beta$ and are BPS
 (they satisfy a first order equation
which implies the second order field equation). For $n>2,$ there are also
BPS domain wall junctions in which three or more domain walls meet
at a point, dividing the plane into sectors with a different
vacuum occuring in each sector. The plane can be tiled by almost-BPS
junctions, in the sense that each junction is locally close to a BPS
junction, but the particular BPS equation varies at different junctions
and some junctions are required to be anti-BPS, in that the orientation
of the junction is reversed.

Returning now to the three-dimensional model defined by the energy
(\ref{energy}) the above discussion suggests that there may be
solutions in which domain wall junctions form a spherical tiling
of the surface of the Q-ball. These domain walls are fully 
three-dimensional, in that although they are localized
close to the Q-ball's surface, they also have a thickness. 
Such domain wall networks will certainly not be BPS, and effects
such as the curvature of the sphere will play an important role,
but the tension formula (\ref{tension}), and other aspects of
planar domain wall junctions, will prove a useful guide
in understanding the results obtained.

\section{Domain Wall Networks}\news\label{sec-results}

\quad\, In this section we describe the results of numerical 
computations to compute static domain wall networks on the surface
 of a Q-ball. Motivated by the suggestions in refs.\cite{BB3,BB4}
we shall concentrate mainly on investigating examples where the domain walls 
lie along the edges of a spherical polyhedron,
 forming junctions at its vertices.

To compute solutions of the static field equations for $\phi,$ 
which follow from the variation of the energy defined by (\ref{energy}),
we solve the associated gradient flow equations,
$\frac{\partial \phi}{\partial \tau}=-\frac{\delta {\cal E}}{\delta \bar\phi}$,
where $\tau$ is an artificial time.
Spatial derivatives
are approximated by second order finite differences and we use a grid
containing $101^3$ points with a lattice spacing $\delta x=0.2.$
As discussed above, the Q-ball surface has a radius $r_*\approx 6$
and the walls have a width of order $v/\beta.$ To make the width
of the walls of order one, which is a reasonable fraction of the diameter
 of the sphere,
we choose $v=\beta,$ where $\beta=0.1$ as before.

The simplest domain wall network consists of just two junctions, located
at antipodal points on the sphere. For the theory with $n$ vacua we may 
take $n$ equally spaced walls, lying along geodesic arcs joining the
two junctions. The domains are equal area segments of the sphere and
each vacuum is attained only in one segment, with neighbouring
segments containing neighbouring vacua, so that all walls have minimal
tension. For later reference, we refer to such a solution as an $n$-segment
network.

\begin{figure}[ht]
\begin{center}
\leavevmode
\epsfxsize=4cm\epsffile{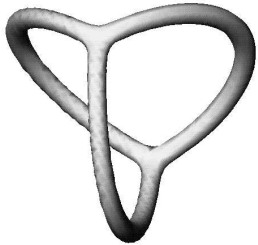}
\caption{Energy density isosurface for the 3-segment network.
}
\label{fig-3seg}
\end{center}
\end{figure}

It is a simple task to create initial conditions which will yield
an $n$-segment network. $\phi$ is taken to be the vacuum  $j,$
where $1\le j\le n$, at the points which satisfy all the following conditions;
(i) $|x_3|<r_*-\mu_1$, (ii) $|r-r_*|<\mu_2$, 
(iii) ${2\pi(j-1)}/{n}+\mu_3<\varphi<{2\pi}{j}/n-\mu_3,$
where $r,\theta,\varphi$ are spherical polar coordinates.
At points which do not satisfy all these conditions then $\phi=0.$
The values $\mu_1=1,\mu_2=2,\mu_3=0.1$ were used, though others
yield the same results.
Condition (i) removes patches around the north and south poles, 
which is where the junctions will form. Condition (ii) seeds a
condensate for $\phi$ only around the surface of the Q-ball.
Condition (iii) sets the correct ordering of vacua, leaving
a small buffer between the segments for the domain walls to form.
In fig.~\ref{fig-3seg} we display a surface of constant energy density
for the solution obtained from the above initial conditions in the
simplest case of $n=3.$ The three curved domain walls are clearly evident
 as the geodesic arcs, and the  3-wall junctions can be seen at
anitpodal points on the sphere. 

\noindent Similar solutions have been computed for other values of $n.$

An important point to clarify is the symmetry of an $n$-segment network.
The energy density has the point group symmetry $D_{nh},$ but of course
the fields themselves do not have this symmetry, since different vacuum
values are attained in each of the $n$ segments. In theories with a
global symmetry, in this case a discrete $\Z_n$ symmetry, the notion
of a field configuration having a spatial symmetry is usually understood
 to mean that the field configuration obtained after acting with the
spatial transformation is equivalent to the original field configuration
upto the action of the global symmetry group. In particular this implies that
the energy density is strictly invariant. However, the $n$-segment network
is not $D_{nh}$ symmetric in this sense, even though the energy density is.
The easiest way to see this is to note that the junction at the north
pole has the opposite orientation to the junction at the south pole -
this is the analogue of a BPS and anti-BPS junction in the planar
Wess-Zumino theory. The action of the global symmetry group can not
reverse the orientation of a junction, so they must be considered as
different objects, even though they have the same energy density.
Thus an $n$-segment network has only a cyclic $C_n$ symmetry.

 A junction in which the phase of the 
vacuum values increases (upto a $2\pi$ periodicity) as each consecutive
domain is entered in an anti-clockwise direction will be termed
a positive junction, whereas if the phase of the vacuum values
continually decreases it will be called a negative junction.
Note that for $n>3$ not all junctions are either positive or negative.
With this terminology we can now state that an $n$-segment network has
one positive and one negative junction.

Even in the planar Wess-Zumino model (\ref{WZ}) a static domain wall
network is unstable \cite{Sa}, which is perhaps not surprising
given that it must consist of both locally BPS and anti-BPS junctions.
Unstable modes are associated with a self-similar shrinking of a 
domain, which in the thin wall limit (where the width of a domain
wall is neglected) become BPS zero modes.
It is therefore expected that domain wall networks on the surface
of a soliton will also be unstable, and this is indeed the case.
The $n$-segment network is unstable to perturbations which break
the cyclic $C_n$ symmetry. For example, if the two junctions are
created at points which are not antipodal, or if the segments are
not of equal area, then some domains shrink and eventually disappear,
leaving the surface dominated by a single vacuum  in the
generic case.

Let us now turn to the consideration of more complicated networks,
with several junctions located at the vertices of a spherical polyhedron
and domain walls lying along its edges.
To form a given polyhedron (or more accurately its spherical version)
an assignment must be made of one of the $n$ vacua to each face of
the polyhedron, in such a way that any two faces which meet at an edge
have different vacua. This is to ensure that domain walls
form along the edges of the polyhedron, but in fact, for reasons
that we demonstrate later, a slightly 
stronger condition must be met, in that no faces which meet at a vertex
can share the same vacuum either -- basically such a vertex 
breaks up as domains with the same vacua merge.

By far the most elegant example is the cube in the $n=3$ theory. 
To form a cube an allocation must be made of one of the three vacuum 
values to each of the six faces of the cube,
respecting the above criteria that all faces which meet must have
different vacua. This is simple to attain, with each vacuum
being assigned twice to opposite faces of the cube.
To provide initial conditions for the numerical computations we apply the
following general scheme, which works for any polyhedron with a given
assignment of vacua to its faces. At the centre of each face
we create a sphere of radius $R,$ inside which $\phi$ takes the
required vacuum value $j,$ and outside these spheres we set $\phi=0.$
The only condition on $R$ is that it must be small enough so that
no two spheres overlap, and for the simulations described in this paper
we set $R=2.$

\begin{figure}[ht]
\begin{center}
\leavevmode
\epsfxsize=5cm\epsffile{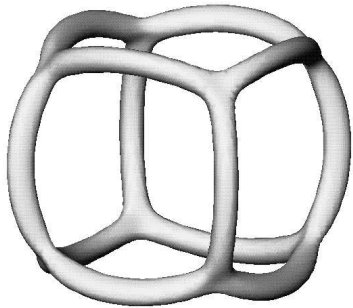}
\caption{Energy density isosurface for the cube network.
}
\label{fig-cube}
\end{center}
\end{figure}
In the case of the cube with the assignment described above,
 the solution computed from the initial
 conditions is presented in fig.~\ref{fig-cube}, by displaying an energy
density isosurface. The energy density is localized on the domain
walls, which lie along the edges of a spherical cube, meeting at
3-wall junctions at the cube's vertices. The energy density has
octahedral symmetry $O_h,$ the symmetry of a cube/octahedron,
because all three species of domain wall have the same tension.
Of the eight junctions, four are positive and lie at the vertices of a
tetrahedron, whereas four are negative and lie at the vertices of the
dual tetrahedron. The configuration is therefore only tetrahedrally
symmetric, even though the energy density has octahedral symmetry.
\begin{figure}[ht]
\begin{center}
\leavevmode
\vskip -3cm
\epsfxsize=15cm\epsffile{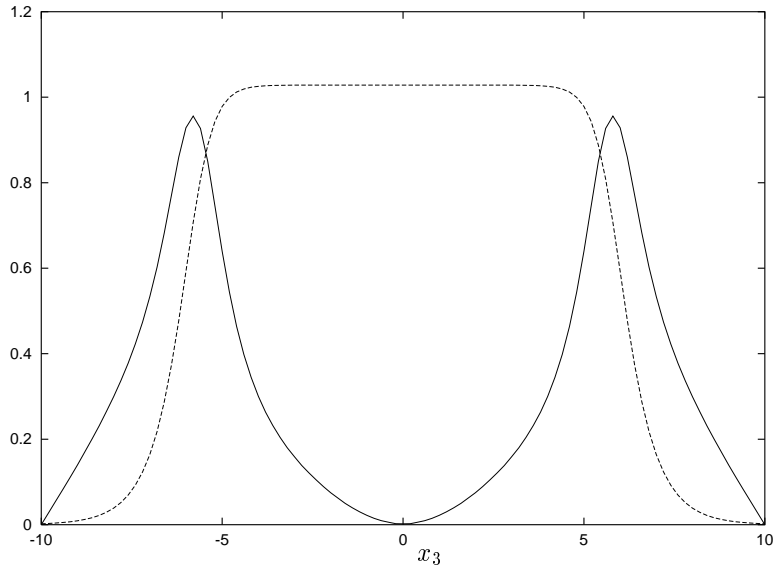}
\vskip -10.5cm
\caption{The functions $|\phi|/v$ (solid curve) and $|\psi|$ (dashed curve)
along the $x_3$-axis, for the cube network.
}
\label{fig-x3}
\end{center}
\end{figure}
In fig.~\ref{fig-x3} we plot the normalized field $|\phi|/v$ and
the magnitude of the Q-ball field $|\psi|,$ along the $x_3$-axis,
for the cube network. This shows how the domain walls are
localized around the surface of the Q-ball, where $|\psi|=\frac{1}{\sqrt{2}}.$

As with all the networks we consider, the cube network has negative modes
in which the area of any given face shrinks. One way to observe such an 
unstable mode is to modify the initial conditions so that one of the
vacuum spheres placed at a face centre has a radius slightly smaller
than the rest. Under the gradient flow evolution this generates a cube 
network with one slightly smaller square face, which shrinks and
eventually disappears, leading to the collapse of the whole network.   

The simplicity of the cube network in the $n=3$ theory might lead
to the expectation that almost any polyhedron can be obtained
as a domain wall network. However, as we shall see below, the cube
network is an exceptional example, being one of the few polyhedra
that can be formed.

\begin{figure}[ht]
\begin{center}
\leavevmode
\epsfxsize=16cm\epsffile{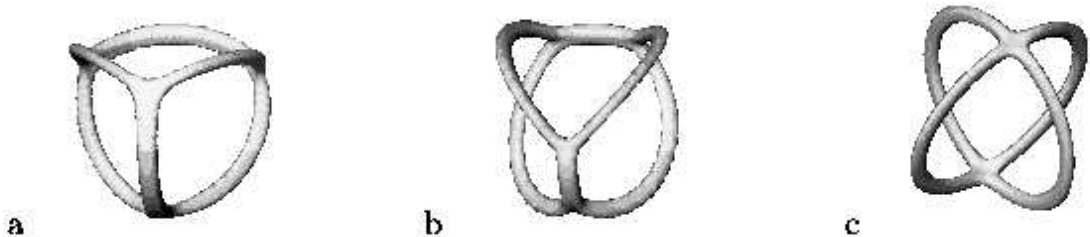}
\caption{Energy density isosurface for an attempted 
tetrahedral network; \ a) near the start of the gradient flow
evolution, b) midway through the evolution, c) at the end of
the evolution.
}
\label{fig-tet}
\end{center}
\end{figure}

To highlight the difficulties in creating a given polyhedron let
us discuss another symmetric example, the tetrahedron. 
Like the cube, the tetrahedron is trivalent, so naively one might
think that it could be formed in the $n=3$ theory, since only
3-wall junctions are required. However, recall that each face of the
polyhedron must be assigned a vacuum in such a way that no
faces with the same vacuum meet. The four faces of a tetrahedron all
meet each other, so this requires at least an $n=4$ theory.
Let us therefore take the $n=4$ theory and assign a different
vacuum to each of the four triangular faces. Recall that the $n=4$
theory has six species of domain walls, four of which we shall refer
to as light walls, and two we call heavy walls, since from
the tension formula (\ref{tension}) the two heavy walls have a tension
which is a factor of $\sqrt{2}$ larger than that of the light walls.
The six edges of the tetrahedron are made from four light walls and two 
heavy walls, since each species of wall occurs precisely once.
The two heavy walls are located at opposite edges, and even the energy
density is not tetrahedrally symmetric, having only a $D_{2d}$ 
symmetry. After a short gradient flow evolution the initial conditions 
yield the almost tetrahedral structure displayed in fig.~\ref{fig-tet}a as
an energy density isosurface. Upon further evolution the two heavy walls
shrink, as displayed in fig.~\ref{fig-tet}b, and eventually disappear
leaving the two antipodal 4-wall junctions presented in fig.~\ref{fig-tet}c.
The end point of the evolution is
 therefore a 4-segment network, rather than the desired tetrahedron network.

As we have explained, the root of the problem lies in the inability to
create a network in which even the energy density has tetrahedral symmetry,
because walls with different tensions must be used. This obstruction 
applies to all attempts to create a tetrahedron network for any value
of $n,$ and similar problems plague attempts to create almost all
polyhedra. For example, the dodecahedron requires the $n=4$ theory
and the four vacua can be assigned symmetrically to three faces each, in a way
that covers the twelve faces so that no two meet with the same vacuum.
However, each pentagonal face is formed from three light walls and
two heavy walls, producing the same kind of problem we have seen for
the attempted tetrahedron network. 

\begin{figure}[ht]
\begin{center}
\leavevmode
\vskip -0cm
\epsfxsize=6cm\epsffile{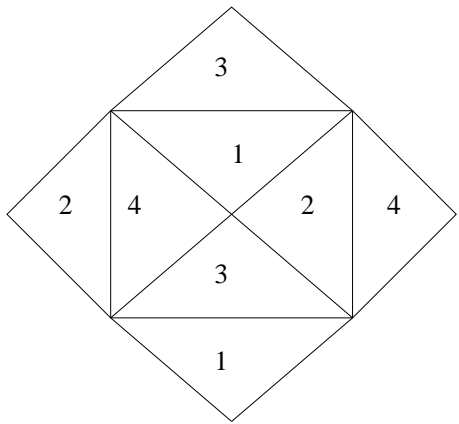}
\vskip -0cm
\caption{A representation of the octahedron together with an assignment
of vacua to the faces.
}
\label{fig-faces_oct}
\end{center}
\end{figure}

There are no polyhedral networks which are as elegant as the cube,
since walls of different tensions must be employed in all other cases.
However, there are a limited number of other polyhedra which can
be formed, in a fashion, by arranging the locations of the heavy
walls in a particularly symmetric way. The simplest example is an
octahedron network in the $n=4$ theory. A representation of the octahedron
is displayed in fig.~\ref{fig-faces_oct} together with an assignment of
one of the four vacua to each face.

\begin{figure}[ht]
\begin{center}
\leavevmode
\epsfxsize=5cm\epsffile{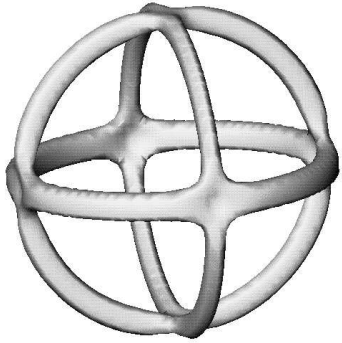}
\caption{Energy density isosurface for the octahedron network.
}
\label{fig-oct}
\end{center}
\end{figure}

As can be seen from fig.~\ref{fig-faces_oct}, each vacuum  is 
assigned twice to opposite faces of the octahedron. The 4-wall
junctions at the north and south poles both contain only light walls,
and this accounts for eight of the edges of the octahedron, but the
remaining four edges are heavy walls. However, these heavy walls
all lie on the equator and have equal length. Although the domain
wall junctions lie at the vertices of an octahedron, and the
walls lie along the edges of a spherical octahedron, the energy density
is only $D_{4h}$ symmetric, because of the different wall tensions.
However, this symmetry is enough to allow both light and heavy
walls to sit in equilibrium, producing an octahedron network.
This is displayed in fig.~\ref{fig-oct} as an energy density
isosurface.
\begin{figure}[ht]
\begin{center}
\leavevmode
\vskip -0cm
\epsfxsize=8cm\epsffile{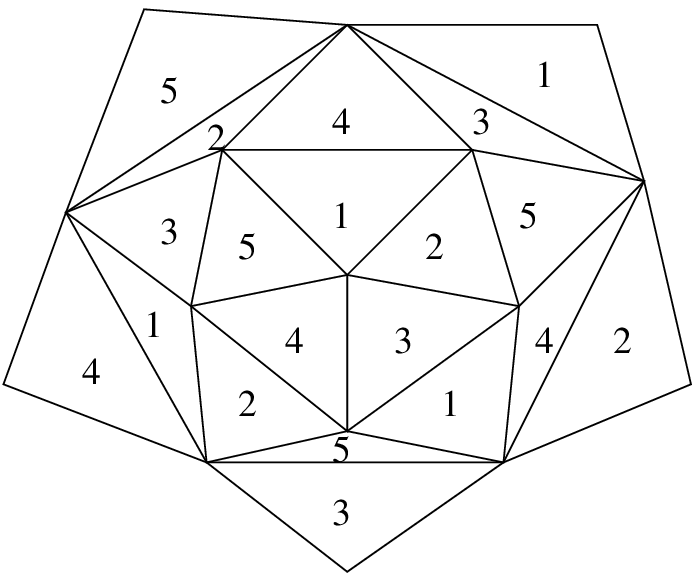}
\vskip -0cm
\caption{A representation of the icosahedron together with an assignment
of vacua to the faces.
}
\label{fig-faces_icos}
\end{center}
\end{figure}
\begin{figure}[ht]
\begin{center}
\leavevmode
\vskip -0cm
\epsfxsize=6cm\epsffile{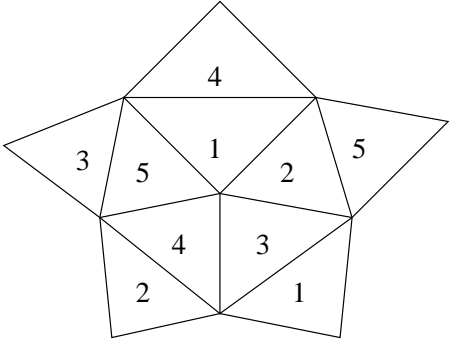}
\vskip -0cm
\caption{A representation of the pentagonal bipyramid
 together with an assignment of vacua to the faces.
}
\label{fig-faces_5bi}
\end{center}
\end{figure}
It might appear that a variety of polyhedra could be formed in a similar
manner as for the octahedron, by a symmetric arrangement of both light and
heavy walls. However, it turns out that even very symmetric arrangements
often fail to form the intended polyhedron.
 As an example, consider the icosahedron,
which requires the $n=5$ theory to form the necessary pentamers. 
There are ten species of walls; five are light and connect vacua whose
phases differ by $\pm 2\pi/5$ and the remaining five heavy walls 
connect vacua with phase differences of $\pm 4\pi/5.$
A very symmetric assignment of the five vacua to the twenty faces
of the icosahedron is presented in fig.~\ref{fig-faces_icos}. 
Each vacuum is assigned four times, there are fifteen light walls
and fifteen heavy walls, arranged with a cyclic $C_5$ symmetry.
However, even this symmetric arrangment does not form an icosahedron,
instead producing a 5-segment network as the end point of the gradient 
flow evolution.

The mechanism for the collapse of the icosahedron can be understood
on a simpler, but related, example. This is the pentagonal bipyramid in the
$n=5$ theory, with the assignment as presented in fig.~\ref{fig-faces_5bi}.

As seen in the figure, the five vacua are assigned to the top pentamer
in a cyclic order, and to the bottom pentamer in the same cyclic order,
when viewed from above. Note that there is a relative rotation of the
bottom pentamer through an angle of $4\pi/5,$ in order to satisfy the 
requirement that no two faces containing the same vacuum meet. Clearly
a rotation through an angle of $6\pi/5$ would also suffice, and
yield a similar configuration. The important observation is that neither
of these two alternatives can place the two faces with a given vacuum
 directly opposite each other, so there is always a preferred direction
of rotation in which to untwist the configuration so that faces with
the same vacuum can align and unite. It is precisely such an untwisting
motion that destroys both the icosahedron and the pentagonal bipyramid.

\begin{figure}[ht]
\begin{center}
\leavevmode
\epsfxsize=14cm\epsffile{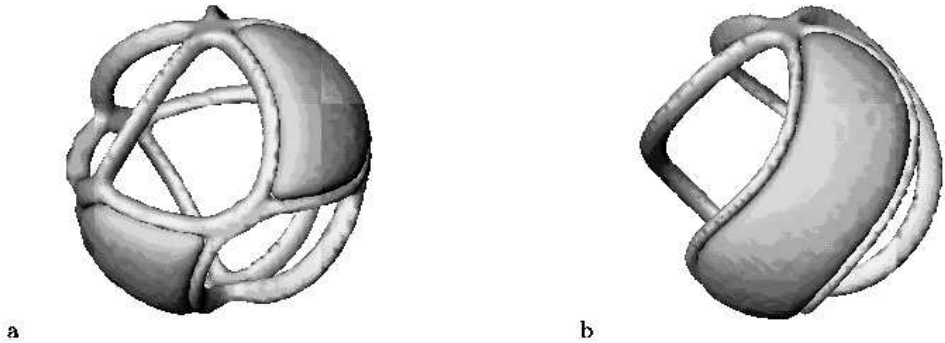}
\caption{Energy density isosurface for an attempted 
pentagonal bipyramid, with the vacuum $\phi=v$ distinguished
 ; \ a) near the start of the gradient flow evolution,
b) midway through the evolution.
}
\label{fig-5bi}
\end{center}
\end{figure}
Fig.~\ref{fig-5bi} displays an energy density isosurface, at two 
different times during the gradient flow evolution, of the 
attempted bipyramid associated with fig.~\ref{fig-faces_5bi}.
Also shown is
an isosurface where $\phi\approx v,$ indicating where this particular
vacuum is attained. The twisting is already clearly visible in 
fig.~\ref{fig-5bi}a, where it can also be seen that 
4-wall junctions have split into two 3-wall junctions 
through the creation of an additional wall. Fig.~\ref{fig-5bi}b shows
the same configuration after further gradient flow evolution, where the
twisting has succeeded in uniting the domains, producing a twisted
version of the 5-segment network. Upon further evolution this 
network untwists and finally forms the 5-segment network.
A similar, though more complicated, picture results from the evolution
of the icosahedron network.

\begin{figure}[ht]
\begin{center}
\leavevmode
\epsfxsize=5cm\epsffile{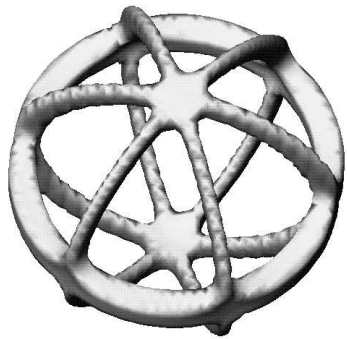}
\caption{Energy density isosurface for the hexagonal bipyramid network.
}
\label{fig-6bi}
\end{center}
\end{figure}
From the above discussion it is clear that even a bipyramid can not
be obtained if each of the antipodal junctions contains an odd number of walls.
However, in the $n$-vacua theory, where $n$ is even, a bypiramid with
two $n$-wall junctions can be formed, because the top junction can be
rotated by $\pi$ relative to the bottom junction, placing the same
vacua at opposite faces, and thus in unstable equilibrium since there
is no preferred direction to begin the untwisting. The simplest example
is $n=4,$ in which case the bipyramid is the octahedron that we have 
already seen. The next example, $n=6,$ is displayed in fig.~\ref{fig-6bi},
where an energy density isosurface is shown to form the 
hexagonal bipyramid at the end of the gradient flow evolution. The energy
density has a $D_{6h}$ symmetry in this example.

\section{Archimedean Networks}\news\label{sec-arch}

\quad\,  In the previous section we have seen that a domain wall network
requires a consistent assignment of vacua to the faces of a polyhedron
and a resulting balance of tensions. This is most easily realized if the
network is symmetric, and since we have already discussed the Platonic
solids the most natural class to consider next are the Archimedean solids.
Rather than compute any examples in detail, we shall apply the criterion
that we have outlined in the previous section to determine the Archimedean
networks that can be obtained in a theory with three vacua.
 
\begin{figure}[ht]
\begin{center}
\leavevmode
\epsfxsize=15cm\epsffile{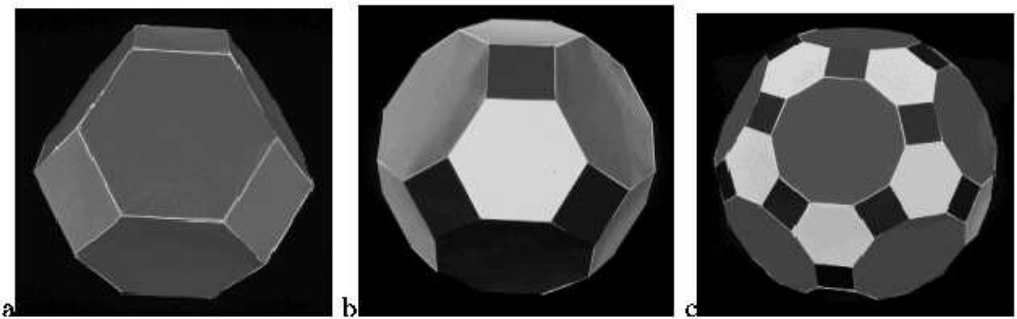}
\caption{The 3-colourable Archimedean solids; a) the truncated octahedron,
 b) the great rhombicuboctahedron, c) the great rhombicosidodecahedron.}
\label{fig-arch}
\end{center}
\end{figure}

The Archimedean solids, of which there are thirteen, are generalizations
of the Platonic solids. In an Archimedean solid all
vertices are equivalent, but the faces are formed from more than one type
of regular polygon. Of the thirteen Archimedean solids it is a simple
task to determine that there are three
which can be 3-coloured in the way required to be a
domain wall network in a theory with three vacua. These three, which
are displayed in fig.~\ref{fig-arch}, are the
truncated octahedron, the great rhombicuboctahedron and the great
rhombicosidodecahedron.
The truncated octahedron has 14 faces, of which 6 are squares
and 8 are hexagons. At each of the 24 vertices one square and two hexagons
meet. The three vacua can be assigned by allocating one vacuum to the
squares and assigning the remaining two vacua alternately to the four
 hexagons around each square.
The great rhombicuboctahedron contains 12 squares, 8 hexagons and 6 octagons,
making a total of 26 faces. At each of the 48 vertices one square, one
hexagon and one octagon meet, so an assigment of a different vacuum to
each of the three types of polygon can be made.
The great rhombicosidodecahedron has 62 faces, consisting of 30 squares,
 20 hexagons and 12 decagons. A square, a hexagon and a decagon meet at
each of the 120 vertices, so again each type of polygon can be allocated
a different vacuum.

None of these Archimedean solids can arise as domain wall networks
in the $n=3$ model, since all wall tensions are equal, 
so only a junction with $120^\circ$ angles exists.
However, as discussed in ref.\cite{BB3}
in a general context, it is possible to deform a theory by breaking
the $\Z_3$ symmetry so that a junction with specified angles exists.
Since all vertices are equivalent in an Archimedean solid, then for each
of the three examples described above it is possible to make a 
(different) symmetry
breaking deformation of the potential so that the three wall tensions
have the correct ratios to form a junction with the angles required for
that particular solid. In the case of the truncated octahedron the
deformation breaks the $\Z_3$ symmetry to a $\Z_2$ symmetry, whereas
for the other two solids the deformed potential will have no discrete
symmetry.

\section{Conclusion}\news\label{sec-conc}

\quad\, Using numerical methods, domain wall networks on the surface
of a Q-ball have been studied in a simple model. A small number
of polyhedral networks have been found, the cube being the most
elegant example, but it has been demonstrated that most polyhedra
will not arise because of severe constraints on assigning vacua to
faces in such a way that wall tensions are in equilibrium. The analysis
presented here reveals that the suggestions of refs.\cite{BB3,BB4} were
too optimistic, and one does not expect to find fullerene-like networks  
of domain walls on the surface of any kind of soliton. This is
because  our
considerations involve only very general arguments, such
as the assignment of vacua and wall tensions. The domain wall
networks that we have been able to construct
 should exist in a variety of different
theories in which the Q-ball soliton is replaced by a topological soliton,
such as a monopole or Skyrmion, with additional fields coupling to the
length of the Higgs field or sigma field respectively.

Finally, it is interesting to note that in the $n=3$ theory the two networks 
constructed in this paper are examples of Lamarle networks.\footnote{I thank
Gary Gibbons for pointing this out to me.} A Lamarle network is a collection
of great circle segments on the sphere which intersect three at a time with
$120^\circ$ angles. There are precisely ten Lamarle networks \cite{La,He}
and it is easy to see that of these ten only two (the 3-segment network
and the cube network) can be 3-coloured in the way required to be a
domain wall network in a theory with three vacua. Lamarle networks play
a role in the study of soap films and bubbles \cite{AT}, which is 
another system in which an equilibrium of tensions is required, but
only triple junctions are possible in this application \cite{Ta}.

\section*{Acknowledgements}
Many thanks to Dionisio Bazeia and Francisco Brito for helpful correspondence.
I acknowledge the EPSRC for an advanced fellowship.\\

\end{document}